# Probing exciton dispersions of freestanding monolayer $WSe_2$ by momentum resolved electron energy loss spectroscopy

Jinhua Hong[1], Ryosuke Senga[1], Thomas Pichler[2], Kazu Suenaga[1,*]

[1]Nanomaterials Research Institute, National Institute of Advanced Industrial Science and Technology (AIST), Tsukuba 305-8565, Japan

[2]Faculty of Physics, University of Vienna, Strudlhofgasse 4, A-1090 Vienna, Austria

E-mail: suenaga-kazu@aist.go.jp

**Momentum resolved electron energy-loss spectroscopy was used to measure the dispersions of excitons in a free-standing monolayer of $WSe_2$. Besides the parabolically dispersed valley excitons, a sub-gap dispersive exciton was observed at specific $q$ values for the first time. The simultaneous STEM-ADF imaging of the local probed areas suggests that this sub-gap exciton is attributed to the prolific Se vacancies.**

Valley[1-3] and defect-bound excitons[4-7] have been attracting great interest in the valleytronic, opto/electronic properties of monolayer transition metal dichalcogenides (TMDs). In these 2D systems, single quantum photon emission[4-6] was recently discovered, whereas its origin remains unclear and controversial. Compared to the valley exciton with predicted dispersions[8-11], defect-induced exciton is much less exploited, while both exciton $q$-$E$ dispersions are of great significance to understand the exciton dynamics. In optical absorption experiments, valley excitons in 2D TMDs were excited only at $q$=0 limit, and therefore the dispersion relationship is inaccessible with the vanishing nonzero-$q$ excitation. Inelastic X-ray/neutral scattering[12] is feasible to obtain the exciton dispersion but measures only bulk materials in millimeter size, and is not capable to measure the monolayer TMDs system undergoing an indirect-to-direct bandgap crossover. On the other hand, reflection EELS always requires a substrate for the sample, introducing non-negligible substrate-sample interaction. Hence, $q$-EELS in transmission[13,14] (mostly in TEM) is a best choice to probe the intrinsic exciton dispersions of freestanding monolayer TMDs. In addition, the $q$-EELS measurement in TEM has superior advantage in spatial resolution, and one can visualize the atomic structures of local areas with possible defects and impurities.

In literature, $q$-EELS has been successfully used to measure the plasmon and phonon dispersions of TMDs[15], graphene[16] and h-BN[17,18]. The dispersions of the basic valley exciton and defect induced exciton (shown in Fig.1d) have never been reported in monolayer TMDs system due to the challenge in detection sensitivity and signal-to-noise ratio, despite their importance to perceive exciton dynamics and derive key parameters such as effective mass, mobility, etc. Here, taking advantage of the state-of-the-art energy/momentum resolution ($\Delta$E~40meV, $\Delta q$=0.025Å$^{-1}$), we employ $q$-EELS to probe the $q$-E dispersions of various excitons in monolayer WSe$_2$ with prolific atomic defects.

Experimental $q$-EEL spectra of freestanding monolayer WSe$_2$ were acquired in the standard diffraction mode in a TEM at 60 kV, as shown in Fig.1a. Spectra were taken from clean monolayer region of good crystalline quality (Fig. 1c and Fig.S1)[19]. With a monochromator, the energy resolution of 40~50 meV can be easily accessible (Fig.S2). In the diffraction space, we used a spectrometer entrance aperture (SEA) to select the specific in-plane momentum $q$ along ΓM and ΓK directions (Fig.1b). The momentum resolution is defined by the size of SEA (1mm ~ 0.2mrad ~ 0.025Å$^{-1}$). In the following sections, the terminology "momentum $q$" always refers to in-plane momentum transfer, since the out-of-plane momentum can be neglected for our $q$ range measured (Fig.S3). In real space, the spatial resolution is determined by the selected area aperture, which corresponds to an area of the monolayer in diameter ~200 nm. Figure 1d is a schematic illustration of electronic transitions from band edges or defect bands, which results in valley exciton "A" and possible defect exciton "$x$", which will be mentioned frequently later.

Figure 2 shows the $q$-E maps obtained along two typical in-plane orientations - ΓM and ΓK directions. For small momentum transfer $q$ = 0~0.03 Å$^{-1}$, four branches of exciton peaks are clearly visible: A at 1.69 eV, B at 2.10 eV (here A, B peaks refer to the existing literatures) and C at 2.50 eV, D at 3.00 eV (labeled as A′, B′ in other optical measurements[20,21]). The former two are often attributed to the intravalley excitons A$_{1s}$ and B$_{1s}$ from spin-splitting band-edge van Hove singularities[22] such as K$_v$-K$_c$ transitions, and the latter peaks C, D from higher order Rydberg excited states like A$_{2s}$ and B$_{2s}$ (or A′, B′) [8,20,23]. Dispersive behaviours are unambiguously observed for the three branches of A, B, and C in Fig.2. As $q$ increases, the three branches of excitons A, B, C present blueshifts with decreasing intensity but different dispersive curvatures. While the lowest-energy exciton A shows an everlasting intensity up to $q$ = 0.2 Å$^{-1}$, the other excitons quickly disappear and get drowned into the background as $q$ increases. It is worthwhile to mention that excitons can only survive in the range of $q$ < 0.2 Å$^{-1}$ in our experimental measurements. The exciton signal of monolayer WSe$_2$ for higher $q$ is quite weak and undetectable with low signal-to-noise ratio.

The raw experimental $q$-EEL spectra are displayed in Fig.3 along ΓM and ΓK directions. Here we mainly consider excitons within the energy range ~ 4 eV of our interest. Higher energy excitation (5~8 eV) involves complicated exciton-plasmon interaction (Fig.S4 and Fig.S5) and their interpretations are not within the scope of this paper. As shown in Fig.3, the vertical dashed lines mark the position of all excitons A, B, C, D and E in the $q\rightarrow0$ limit we observed. Above the well-known spin-splitting A, B excitons, the sharp C, D peaks are from Rydberg-state exciton A′ and B′ [8,20].

Along with the decreasing peak intensity, the blueshift of A exciton increases more and more obviously as $q$ increases, indicating a nonlinear increasing dispersion. Compared to sharp A exciton, the next three peaks B, C, D decrease and disappear synchronously on the background of the pre-tail of broad peak E which becomes dominating at $q > 0.11$ Å$^{-1}$ (brown curve in Fig.3a). This background effect is more prominent when the thickness of WSe$_2$ increases (Fig. S6). In MoS$_2$, MoSe$_2$ and WS$_2$ system, the broad C peak due to band nesting has a much larger linewidth $\Delta E > 0.7$ eV than the sharp A exciton with $\Delta E < 0.1$ eV (ref. [23]) and this broad peak persists into high $q$ (Fig.S7a,b). Hence it is reasonable to assign the broad and intense peak E in WSe$_2$ as the electronic transition resulted from band nesting between Γ and Q point (Fig.1b)[22,24]. Strong excitonic effect accounts for the sharp and intense B, C, D exciton peaks before the broad peak E at $q\rightarrow0$. However, a recent $k\cdot p$ model calculation[25] suggests that the broad E peak may also come from other van Hove singularities like the saddle point M (Fig.1b) in the optical band structure. At lower energy end in Fig.3, we found a subgap exciton peak at 1.3 ~ 1.4 eV, as highlighted by the blue arrows and labeled as "$x$". This weak feature can also be seen in Fig.2 and also in other TMDs (Fig.S7c). This suggests that these features may be highly likely induced by defects.

Profiling the peak positions of excitons using Voigt function fitting (Fig.S8), we derive their $q$ dependence in Fig. 4. Here we focus only on A, E, and "$x$" excitons, as B,C,D excitons are on the pre-tail of E exciton and disappear too quickly. In Fig.4a, the measured $q$-$E$ dispersions of A exciton show hardly any in-plane ΓM/ΓK anisotropy. Its dispersion can be well fitted by quadratic function, as expected from its valley exciton nature. Given that A exciton follows $E = \hbar^2q^2/2m^*$, we can derive the effective mass of the A exciton $m^* = 0.65m_e$ (in Fig.4a), which agrees well with the GW calculated $0.72m_e$ in WSe$_2$ monolayer[9]. The value is also comparable with the effective mass of 3D traditional semiconductors.

To correlate with the quasiparticle band structure, we compare our results with theoretical dispersions derived from other band structure calculations, as shown in Fig.4b. In a simplified way (Fig.S9), exciton dispersion can be extracted from the electronic band structure: $E_g(q) = E_c(q) - E_v(0)$, where the initial state is fixed at K$_v$ at $E_v(0)$ and final state at conduction band edge with momentum transfer $q$ and energy $E_c(q)$. This simplified model has been employed in the case of bulk phosphorus of ref.14. More accurately, the initial state can be from any points nearby $K_v$, but with a momentum difference of $q$ with respect to the final state on the conduction band edge. As shown in Fig.S9 and its caption, the parabolic dispersion approximation of valence/conduction bands still yields a parabolic exciton dispersion. Here, the A exciton dispersion are extracted from conduction band edge calculated by Wang et al[26], at the $GW_0$ level including spin-orbit coupling perturbatively. This $GW_0$ dispersion in Fig.4b presents a parabolic relation within the experimental $q$ range we measured. In TMDs, the sharp A exciton peak at $E_A$ is located below the threshold energy $E_g$ (quasiparticle band gap) of the continuum absorption, and their difference is defined as exciton binding energy $E_b = E_g - E_A$ [8,26]. The experimental $q$-$E$ relation is almost parallel with the $GW_0$-derived dispersion, and their difference means a binding energy $E_b = 650$ meV which is almost independent of $q$. This indicates the dispersion-less nature of exciton binding energy of 2D TMDs. Similarly, we derive the binding energy of B exciton as 760 meV. Compared to $GW_0$ dispersion, GW-BSE calculation[9] gives better accuracy in energy, but with a linear dispersion. The dispersion-less binding energy $E_b \sim 0.6$-$0.7$eV, is much larger than that of traditional 3D semiconductors, as a result of less screening effect of the Coulomb interaction of excitons in freestanding monolayer TMDs. This also indicates the Frenkel nature of excitons in TMDs with a Bohr radius much smaller than nearly-free excitons in 3D GaAs, etc., implying a relatively lower charge carrier mobility.

Figure.4c shows the dispersion of defect induced exciton "$x$" and the broad peak E, beyond the range for accurate theoretical prediction. Out of one's expectation, the defect induced exciton is quite dispersive with the increase of $q$. Here, at the low $q$ end ($q = 0$~$0.08$ Å$^{-1}$), the absence of $x$ peak may be ascribed as the strong elastic line hiding the weak signal on its tail. This sub-gap exciton indicates a linearly dispersed defect band in the bandgap of the electronic band structure, within the $q$ range we measured. The right inset of Fig.4c shows the ADF-STEM imaging of monolayer $WSe_2$. And the red circles highlight the most common intrinsic Se vacancies, which most likely account for the defect exciton "$x$". On the other hand, as shown in the left inset of Fig.4c, exciton E presents a more complicated dispersion relation. And it remains to be uncovered whether band nesting or van Hove singularity yields this broad and intense peak.

Besides the exciton dispersion, the $q$ dependence of the intensity of excitons often suggests the transition nature: dipole or multipole transition[27,28]. Among all excitons, we tracked the intensities of A and E excitons. As shown in Fig.5a, we compare the measured $q$-EELS intensity of A exciton with the GW-BSE[9] calculated and analytically-derived dipole approximation[27] results. Here, $q$-EELS, GW-BSE and dipole approximation present a decay of the intensity with the increase of momentum $q$. However, discrepancy appears in the decaying tail (marked by the arrows in Fig.5a) as $q$ further increases > 0.08 Å$^{-1}$. The peak intensity of E presents a much slower decay than the dipole approximation at high $q$, shown in Fig. 5b. The discrepancy at high $q$ imply the significant contribution of non-dipole, eg, quadrupole (Fig. 5b) or higher-order multipole transitions (Fig.S10)[27]. Here, the absence of data points as $q \rightarrow 0$ is to avoid the singularity in the scattering cross section (see Fig.S11). This non-dipole contribution may be responsible for the deviation of $q$-EELS/GW-BSE dispersions.

In contrast with dipole selection rule, quadruple momentum operator ($r^2$) selects the initial and final states with the same parity. Here we simply use LUMO-HOMO orbitals to interpret the possible origin of the dipole-multipole crossover as $q$ increases. For A exciton, critical points $K_v$ ($d_{xy}$, $d_{x2-y2}$) and $K_c$ ($d_{z2}$) are both of even parity, and E exciton are of ($d_{z2}$, $d_{xy}$, $d_{x2-y2}$) orbitals[25]. Hence quadruple transition will get reasonably enhanced for both A and E excitons as in-plane momentum $q$ increases.

Until now, we observe only $K_v \rightarrow K_c$ intravalley excitons (A,B,C,D) or other non-K transition (E) with a limited $q$ (~0.2Å$^{-1}$). Recent GW-BSE calculation[9] of MX$_2$-TMDs (M=Mo,W; X=S,Se) predicts oscillator strength of excitons will also get maximized in high $q$ range for $K_v \rightarrow Q_c$ and $K_v \rightarrow K_c'$ intervalley transitions. Therefore, the further measurements of intravalley and intervalley exciton dispersions by $q$-EELS would be of great interest. However, the experimental intensity for intervalley exciton is extremely low. Because inelastic scattering cross section decrease drastically at high $q$, and it is impossible to get a practical signal-to-noise ratio.

In summary, we used $q$-EELS in TEM to uncover the dispersions of valley and defect excitons of monolayer WSe$_2$. The A exciton present a parabolic dispersion, and its binding energy of 0.65 eV is independent of momentum $q$. The oscillator strength evolution indicates the effects of non-dipole transition on A, E peak at large $q$, which may interpret the discrepancy of $q$-EELS/GW-BSE dispersions. Our work provides an experimental paradigm to detect the exciton dispersion of freestanding monolayer TMDs, which will inspire further research on exciton manipulation in the optoelectronic devices.

We acknowledge the support from JST Research Acceleration Programme and JSPS-KAKENHI (JP16H06333 and JP17H04797). We thank Dr. Quek Su Ying for help in the interpretation of experimental results and theoretical calculations.

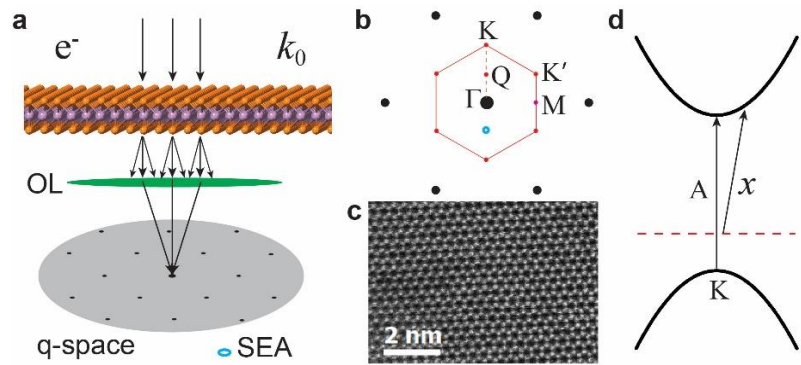

**FIG. 1.** Scattering geometry for $q$-EELS. (a) $q$ space in parallel-beam electron diffraction. The momentum resolution is determined by the post-screen spectrometer entrance aperture (SEA). (b) First Brillouin zone critical points in the diffraction pattern, where the blue circle stands for SEA and selects the specific $q$. (c) Atomically resolved ADF-STEM image of freestanding monolayer WSe$_2$. (d) Schematic illustration of intrinsic band edges at K point and defect band, which yield $K_v \rightarrow K_c$ valley exciton "A" and defect exciton "$x$".

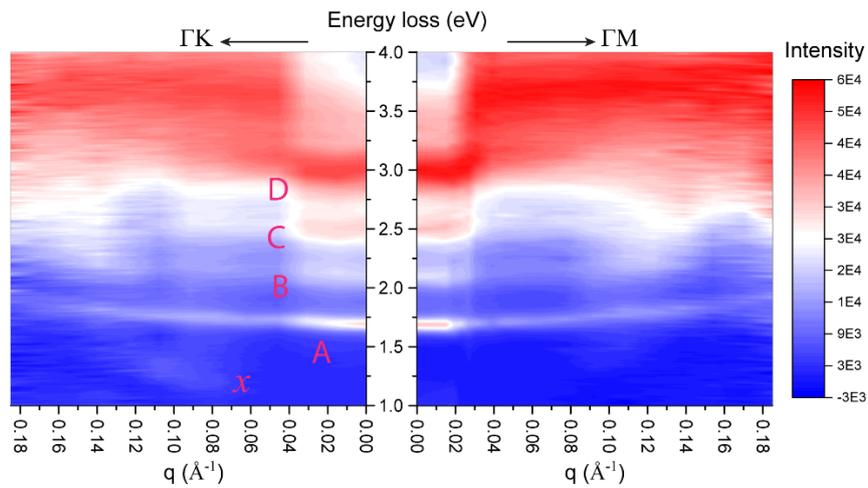

**FIG. 2.** Experimental *q*-E diagram of freestanding monolayer WSe$_2$. The *q*-serial low loss spectra were acquired along ΓK and ΓM directions, respectively. Dispersive bands can be observed: A exciton at 1.70 eV, B exciton at 2.1 eV, and C exciton at 2.5 eV.

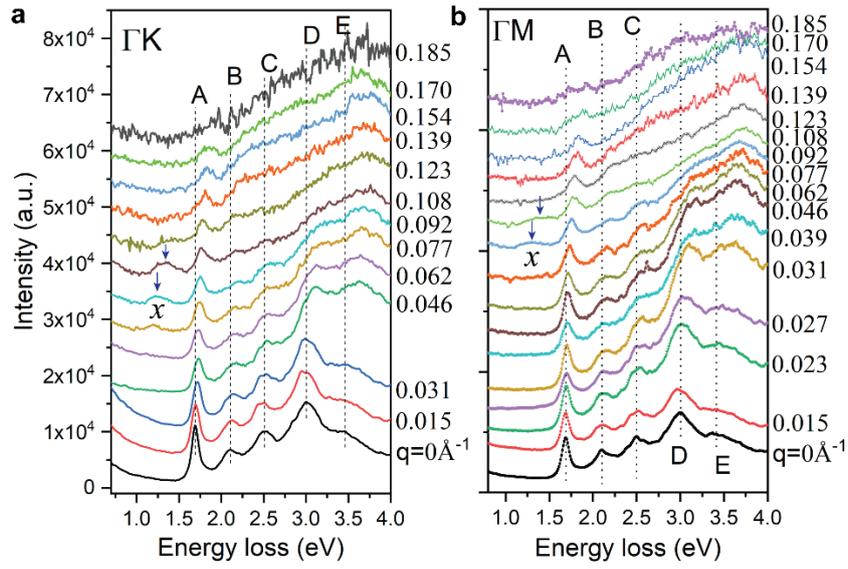

**FIG. 3.** The *q*-dependent EEL spectra fine structures. Dashed lines mark the exciton peaks (A, B, C, D, E) at *q*=0 and guide eyes for the blue-shifting. Besides these major features, a sub-gap defect exciton "*x*" at 1.4 eV emerges and gets enhanced at $q = 0.1$ Å$^{-1}$, highlighted by blue arrows.

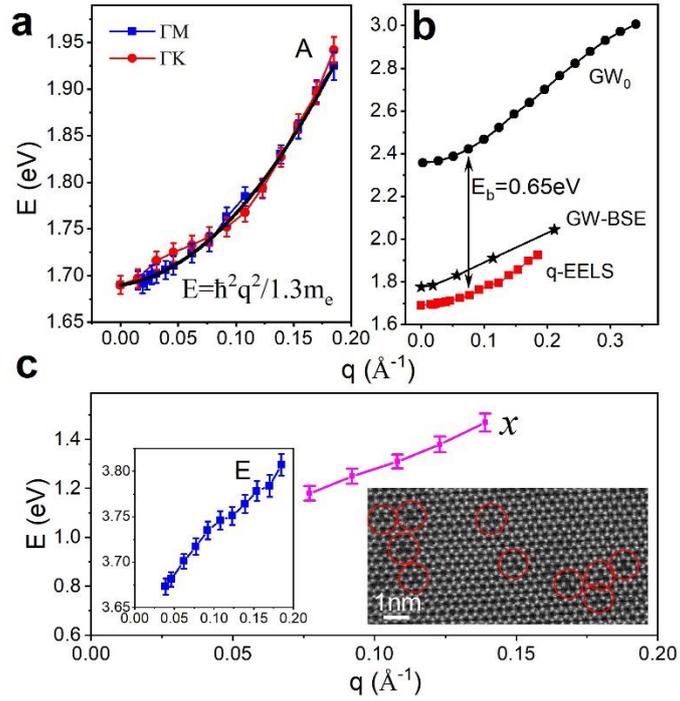

**FIG. 4.** Exciton dispersions. (a) The extracted $q$ dependence of A exciton energy. No obvious in-plane anisotropy for ΓM and ΓK is observed. (b) A comparison of A exciton dispersions by $q$-EELS and other theoretical calculations. $GW_0$ data is extracted from electronic band structure by Wang et al.[26] (Fig.S9) and GW-BSE by Deilmann et al.[9]. (c) Linear dispersion of the defect exciton "$x$", resulted from rich Se vacancies shown in the right inset. The left inset shows exciton E has a complicated dispersion behavior.

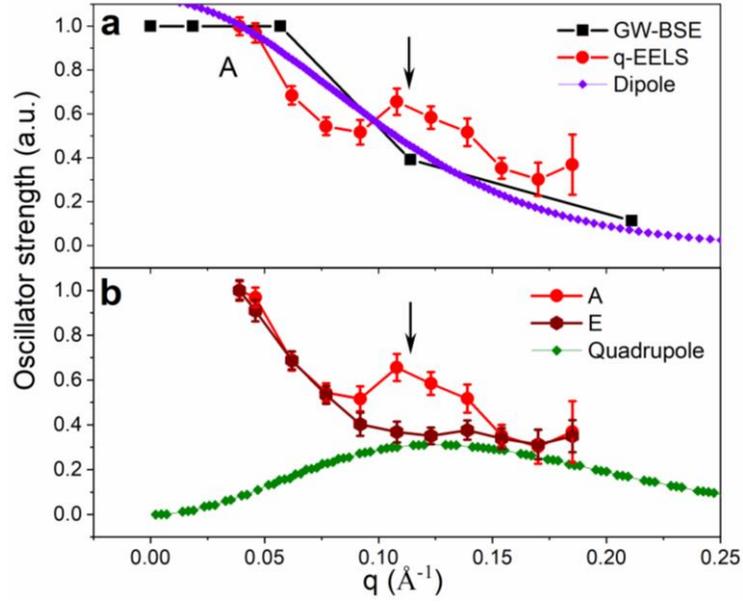

**FIG. 5.** The *q* dependence of exciton peak intensity. (a) The *q* dependence of A excitons. The GW-BSE intensity evolution in black is extracted from the calculation by Deilmann et al[9]. The curve in purple is from dipole approximation (analytical) of the transition matrix element by Knupfer et al[27], where the Bohr radius of A exciton is set as 17 Å (close to the report 10 ~ 20 Å by Stier et al[28] and Berkelbach et al[29]). The black arrow indicates the discrepancy of the decaying tail of the experimental and theoretical *q* distributions. (b) The *q* dependence of oscillator strength of A and E peaks and theoretical quadruple contribution. The quadruple contribution may well interpret the discrepancies of the decaying tails.

# Supplemental Material for Probing exciton dispersions of freestanding monolayer WSe$_2$ by momentum resolved electron energy loss spectroscopy

Jinhua Hong[1], Ryosuke Senga[1], Thomas Pichler[2], Kazu Suenaga[1,*]

[1]Nanomaterials Research Institute, National Institute of Advanced Industrial Science and Technology (AIST), Tsukuba 305-8565, Japan

[2]Faculty of Physics, University of Vienna, Strudlhofgasse 4, A-1090 Vienna, Austria

E-mail:  suenaga-kazu@aist.go.jp

## Methods

**Sample preparation.** Monolayer WSe$_2$ was grown on the silicon substrate (with a 300 nm-thick SiO$_2$ capping layer) by chemical vapor deposition after reduction of WO$_3$ by Se powder in the furnace. The morphology of the monolayer sample is mostly triangle with a size > 50 μm. As-grown WSe$_2$ monolayers were transferred onto TEM grid (with holey carbon) through the standard wet-chemistry method using PMMA. Monolayer sample was annealed at 300°C for 1h in vacuum before EELS measurement to avoid contamination.

**$q$-EELS measurement.** All EEL spectra were acquired in standard diffraction mode in JEOL-TEM-3C2 using parallel beam to illuminate the monolayer sample. With a double Wien filter monochromator, an energy resolution of 40~50 meV can be easily obtained. To balance the signal-to-noise ratio and energy resolution, we chose a proper energy selection slit which gave $\Delta E$ ~ 40 meV. The momentum resolution is determined by the spectrometer entrance aperture and estimated to be ~ 0.025 Å$^{-1}$, which offer a better momentum resolution than the STEM mode employed in recent $q$-EELS measurements. This STEM-EELS offers a momentum resolution on the order of 0.1 ~ 0.2 Å$^{-1}$, meanwhile introducing more severe beam damage on TMDs monolayers even under low voltage. Compared to STEM-EELS, the traditional diffraction EELS is a more suitable choice for the measurement of the $q$-sensitive exciton dispersions of monolayer TMDs. Its good momentum resolution is key to uncovering the exciton band structure since excitons are observed to vanish at $q \geq 0.2$ Å$^{-1}$. The $q$-serial low-loss spectra of our interest (0.5 ~ 10 eV) were collected in Dual-EELS mode, where zero loss peak (or elastic line) can be used for drift correction and the exciton peaks can be determined within a stability or accuracy of ±0.01 eV. For the spectra acquisition, energy dispersion of 0.005 and 0.01 eV/channel and dwell time 0.5 ~ 50 s were used, and it took over 40 min for each spectrum at high $q$. We moved to fresh sample regions after collecting one high-$q$ spectrum, to minimize the effect of beam damage on the excitonic properties.

**STEM imaging.** ADF-STEM images were obtained from JEOL-JEM2100F (3C1) which was operated at 60 kV and equipped with a cold field emission gun and delta corrector. A convergence angle of 35 mrad and acceptance angle of 62 mrad were employed for the ADF imaging.

**Data processing.** To extract the A exciton intensity, the zero-loss peak tail of the acquired low loss spectra was removed by power-law decay function whose window (0.65 ~ 1.0 eV) was fixed for each $q$. To extract the peak energy and oscillator strength non-subjectively, peak fitting using Voigt functions were employed (Fig.S8). All the extracted integrated exciton intensity for each $q$ were normalized according to the Bethe sum rules $\int \omega \cdot \text{Im}\left(-\frac{1}{\varepsilon}\right) d\omega = \pi \omega_p^2 / 2 = \pi n e^2 / 2 m_e \varepsilon_0$, where $\text{Im}\left(-\frac{1}{\varepsilon}\right)$ is the loss function and $\hbar\omega$ is the energy loss. For the safety, we also normalized the integrated intensity of A, E peaks by the zero loss peak for each $q$ (Fig.S12).

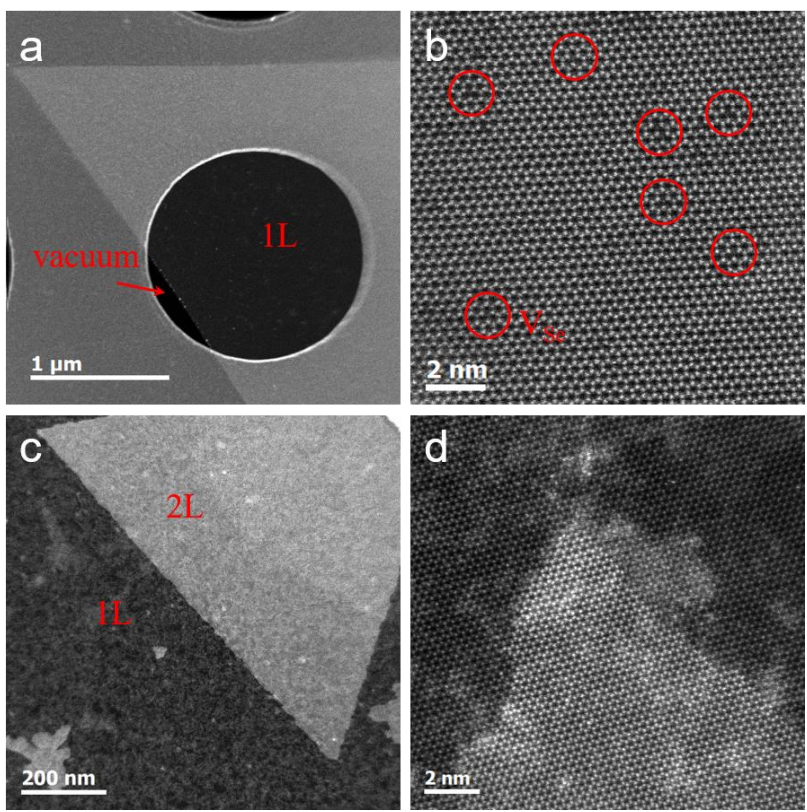

Figure S1. (a, b) Low-magnification and atomically resolved ADF images of monolayer $WSe_2$. Intrinsic Se vacancies can be clearly identified. (c, d) low-magnification and atomically resolved images of bilayer $WSe_2$. The second layer triangle is mostly coherent with the bottom monolayer without rotation angle.

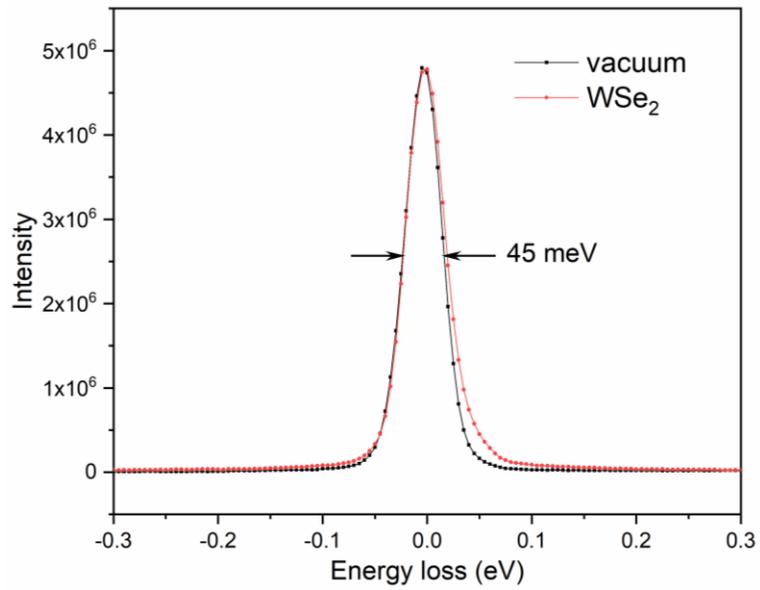

Figure S2. The full width at half maximum (FWHM) of the zero loss peaks of EEL spectra taken from vacuum and monolayer WSe$_2$. The FWHM is slightly broadened on the monolayer.

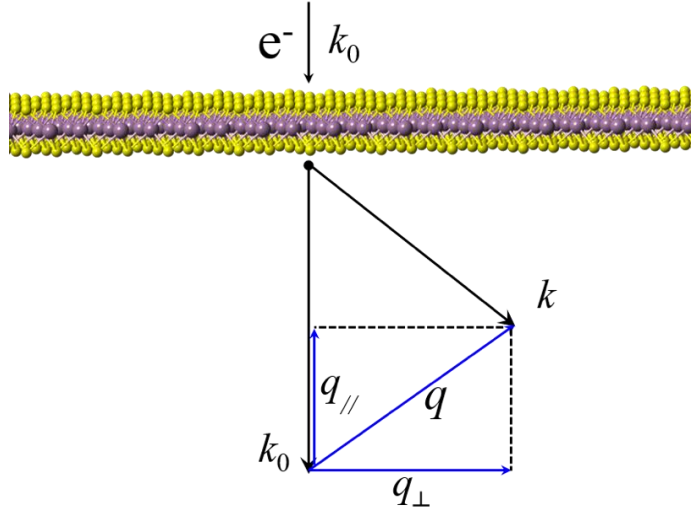

Figure S3. Momentum transfer of the fast incident electron. The momentum transfer vector $q$ can be decomposed into in-plane $q_\perp$ and out-of-plane $q_{//}$. For incident electron with $E_0$=60keV and $k_0$ =128.9Å$^{-1}$, $q_{//}$ can be estimated as $q_{//} = k_0 \cdot \theta_E = k_0 E/(2E_0) = 0.002$Å$^{-1}$ for the energy range $E$~2eV of our interest, which is even much smaller than the $q_\perp$ resolution. Hence this out-of-plane component can be neglected[1] for our $q$ range measured, and $q_\perp$ will be simplified as $q$. Note: to quantify out-of-plane component, one method is to tilt the monolayer to a large angle~70°, which is however far beyond the tilting capability of our narrow pole-piece gap of the electron microscope; another way is maybe to use a convergent beam with $\alpha \sim \theta_E$ =0.02mrad, while such a small angle is difficult to achieve even in TEM mode and the separation of in-plane and out-of-plane contribution is still complicated because of the convolution effect between incident convergent beam and inelastic scattering angular distribution of the out-of-plane contribution.

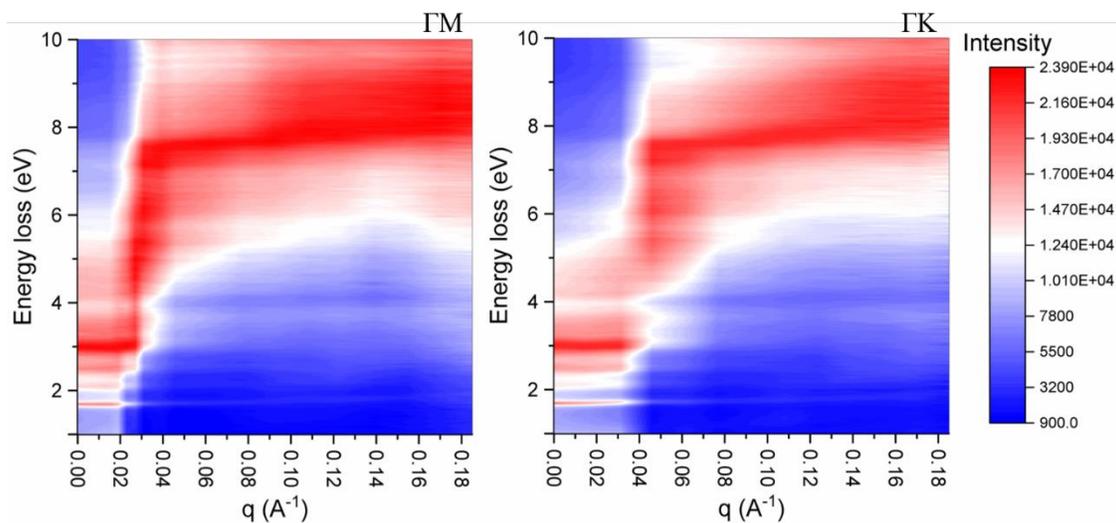

Figure S4. The experimental *q*-E diagrams at higher energy loss along ΓM and ΓK directions.

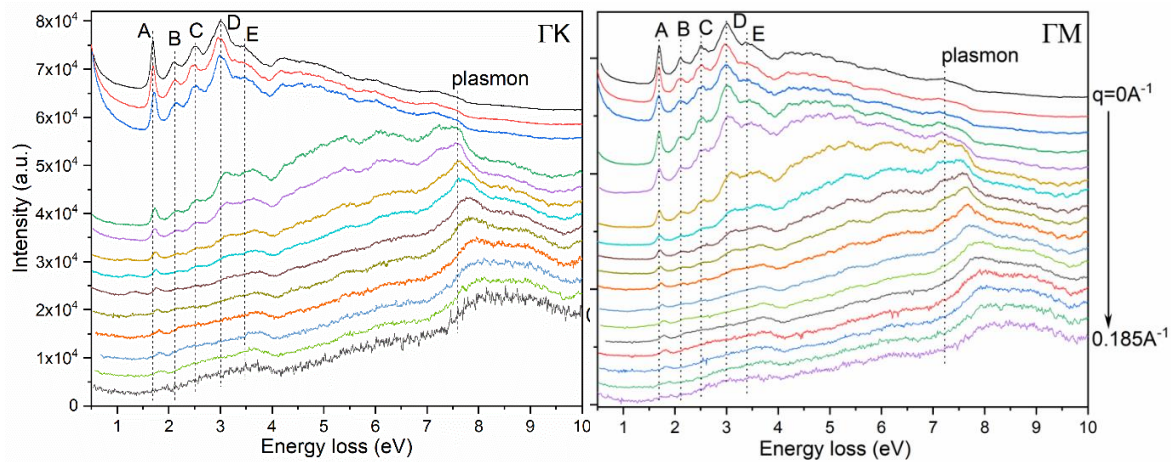

Figure S5. The evolution of higher energy excitation (>4eV) with the increasing *q*.

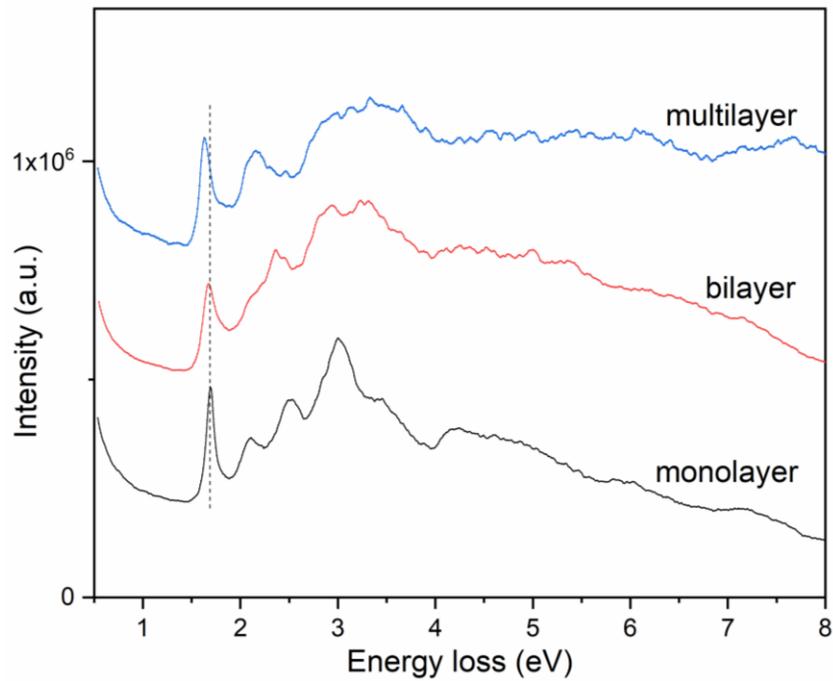

Figure S6. Thickness dependence of the low loss spectra of $WSe_2$ at $q=0$. Exciton peaks get broadened and the pre-background of the E peak at 3.5eV get enhanced when the thickness increases. The exciton peaks get broadened obviously when thickness increases, due to the enhanced phonon scattering. Among these systems of different thickness, monolayer provides the best platform to explore the exciton physics.

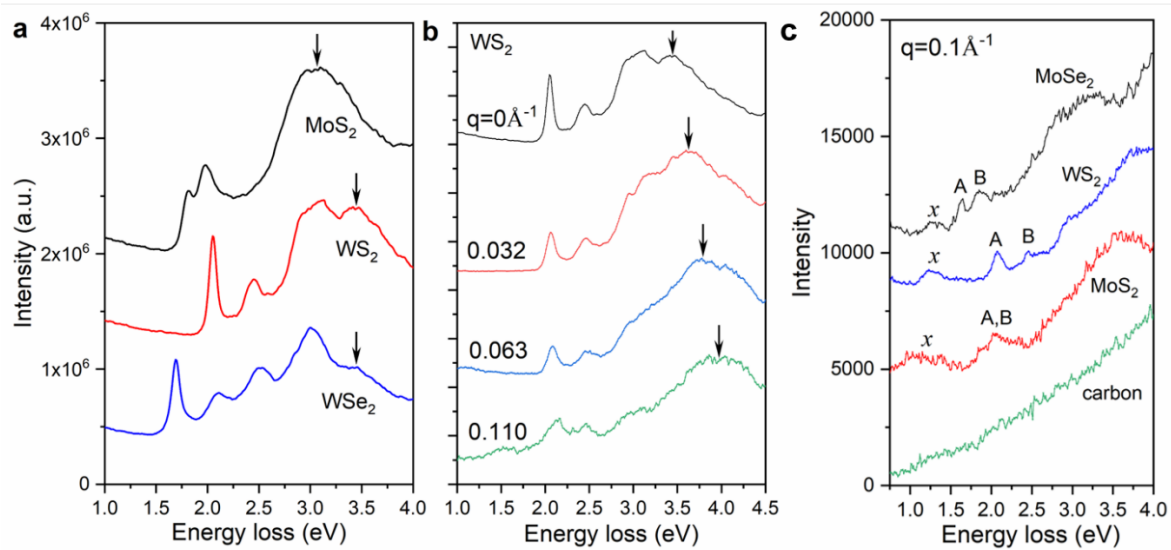

Figure S7. (a) The low loss fine structure of monolayer TMDs at $q=0$Å$^{-1}$. Note the broad and intense peaks marked by black arrows are originated from the same origin - band nesting, and various valley excitons lie on its tail in the lower energy end. (b) The $q$ dependent exciton structures of monolayer WS$_2$, where the band nesting induced intense and broad peaks (marked by arrows) persists into high $q$. (c) The low loss spectra of different materials at intermediate $q=0.1$Å$^{-1}$. Compared with carbon film, the sub-gap signal "$x$" appears in all TMDs, indicating this feature is originated from the material itself and highly likely from defects (strictly, other factors such as lattice strain or surface adsorbates cannot be eliminated).

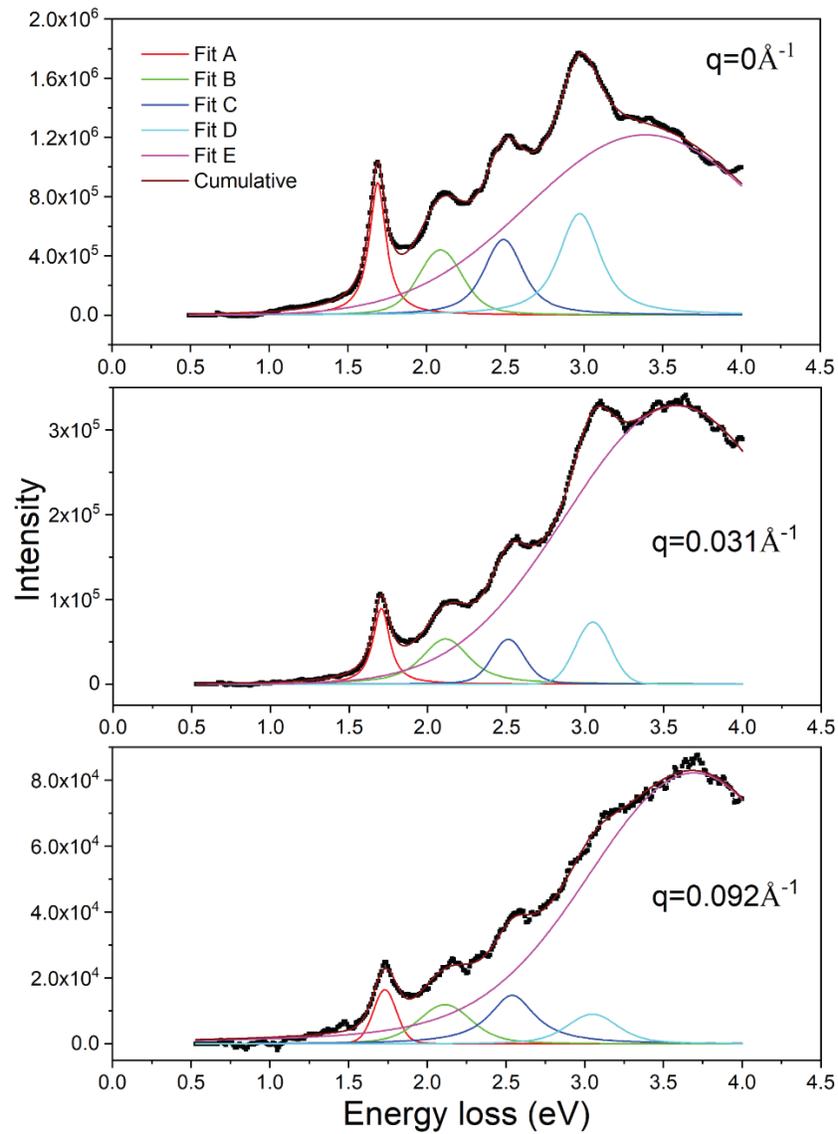

Figure S8. Peak fitting using Voigt function to extract the peak position and integral intensity. Zero loss peak (ZLP) was removed before the curve fitting, where power decay law and fixed window (0.65-1.0eV) was used to remove the ZLP-tail background.

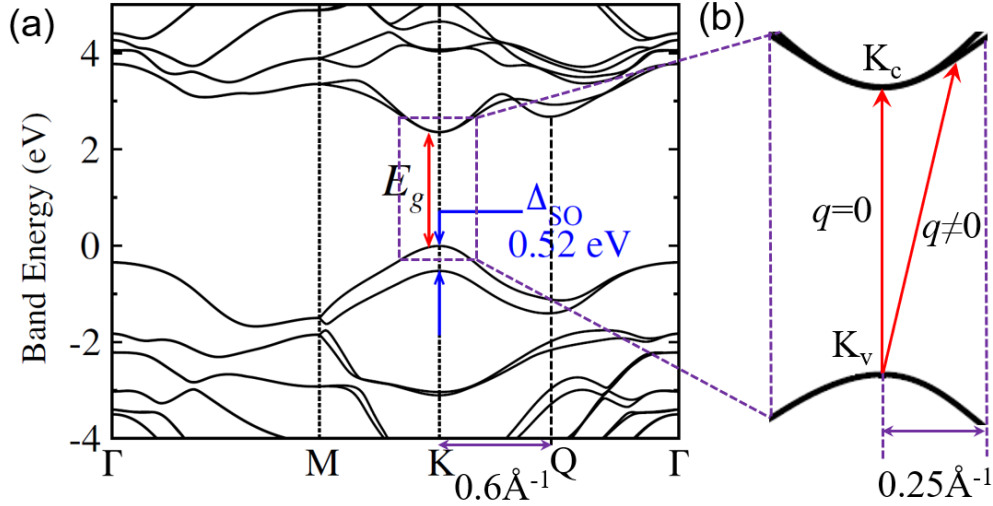

Figure S9. The GW$_0$ band structure and exciton dispersion. (a) GW$_0$ calculated electronic band structure of monolayer WSe2, with spin-orbit coupling included [2]. (b) Illustration of electronic transition with momentum transfer $q$. Note that $|KQ|=0.6$Å$^{-1}$ and the parabolic approximation of bands nearby K point highlighted by dashed rectangle in purple is valid only for $q<0.25$Å$^{-1}$.

The GW$_0$ exciton dispersion can be extracted as: $E_g(q)=E_c(K+q)-E_v(K)$, where the initial state is simplified as $K_v$ point, as used in ref. 14 (PRL 115, 026404, 2015). Then the exciton dispersion is actually the dispersion of conduction band edge. More strictly, the initial state can be from any points nearby $K_v$, but with a momentum difference of $q$ with respect to the final state on conduction band edge. We set the dispersion of valence and conduction bands nearby K point as: $E_c=a_1k^2+E_g(0)$, $E_v=-a_2k^2$, where $a_1$ and $a_2$ are positive parameters. The charge carrier occupation probability is determined by the Boltzmann distribution $f(E_c)=\exp(-(E_c-E_F)/k_BT)$ and $f(E_v)=\exp(-(E_F-E_v)/k_BT)$, respectively. Then the dispersion of exciton with momentum transfer $q$ is $E_g(q) = \int f[E_c(k+q)] \cdot f[E_v(k)] \cdot [E_c(k+q)-E_v(k)]dk = \int [a_1(k+q)^2+a_2k^2+E_g(0)]\exp(-[a_1(k+q)^2+a_2k^2+E_g(0)]/k_BT)dk$. If we set $x=k+a_1q/(a_1+a_2)$, $y_0=E_g(0)+a_1a_2q^2/(a_1+a_2)$, then $E_g(q) = \int[(a_1+a_2)x^2+y_0]\exp(-[(a_1+a_2)x^2+y_0]/k_BT)dx = C_1\exp(-y_0/k_BT)+C_2y_0\exp(-y_0/k_BT)$. Then $E_g(q) = [C_1+C_2E_g(0)+C_2a_1a_2q^2/(a_1+a_2)]\exp\{-[E_g(0)+a_1a_2q^2/(a_1+a_2)]/k_BT\}$. In our $q$ range measured, $a_1a_2q^2/(a_1+a_2) \leq 0.2$eV$\ll E_g(0)=2.4$eV, then $E_g(q) \sim C_3+C_4 \cdot q^2$. This means the exciton dispersion is still parabolic. In Fig.4b, we employ the simplified way (used in ref.14) to derive the exciton dispersion, which is actually the dispersion of conduction band nearby $K_c$.

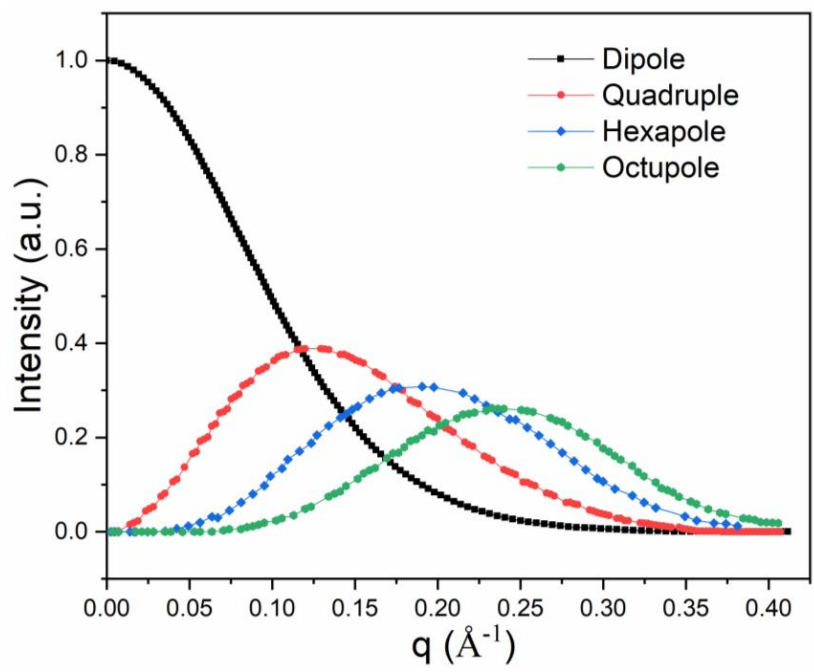

Figure S10. The theoretical *q* dependence of the contribution of various multipole transitions[3].

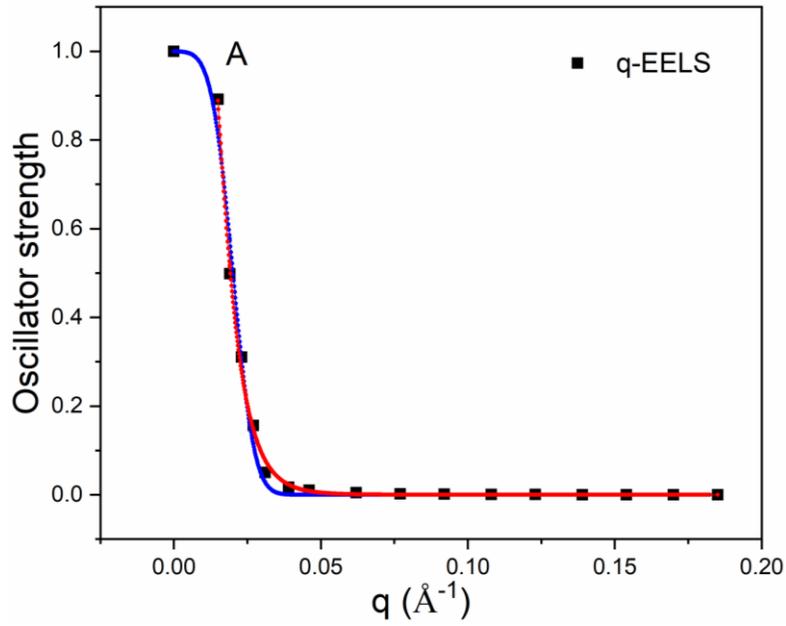

Figure S11. The $q$ dependence of oscillator strength of A exciton, where the $q\to 0$ is also included. Due to the singularity in the scattering cross section $\frac{d^2\sigma}{dqdE} \propto \sigma_{Ruth} \cdot \frac{\text{Im}(-1/\varepsilon)}{q^2}$ when $q\to 0$, the finite energy and momentum resolution leads to an abrupt quasi-elastic ZLP tail which follows a decaying $q$ dependence as highlighted by the good match to the red and blue fitting curves using lognormal distribution function (Lognorm) and normal cumulative distribution function (NormCDF), respectively.

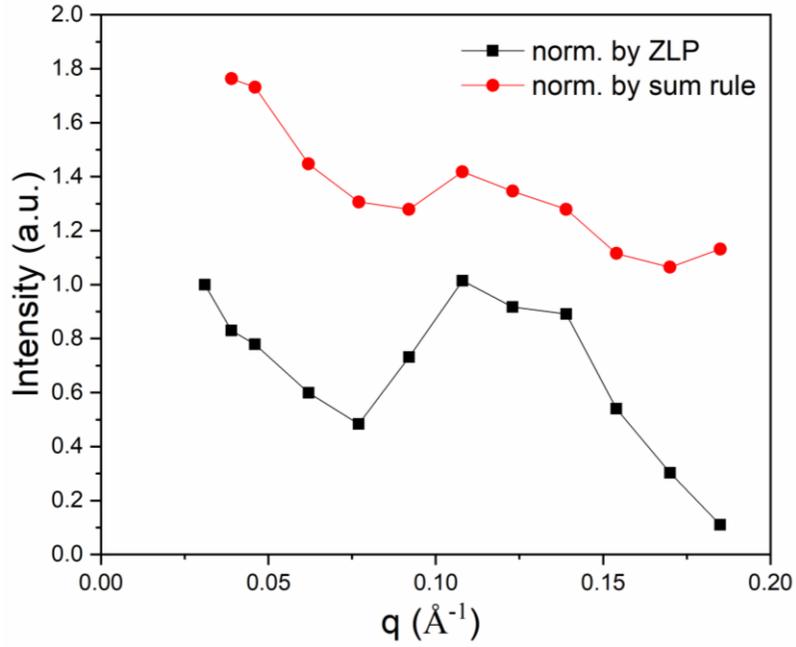

Figure S12. The *q* dependence of the oscillator strength of A exciton. The intensity of A exciton for each spectrum can be normalized in two ways: one is to be normalized by the elastic line zero loss peak (ZLP); another is by the sum rule. In the former case, the peak area of A exciton per unit acquisition time is normalized by the ZLP intensity over unit acquisition time for each *q*, then we obtain the curve in black dots. In the latter case, spectra at each q are rescaled according to the Bethe sum rule - $\int_0^\infty \omega \cdot \mathrm{Im}(-1/\varepsilon)\, d\omega = \frac{\pi \omega_p^2}{2} = \frac{\pi n e^2}{2 m_e \varepsilon_0}$, and then their integral intensity is normalized and presented as the curve in red dots. Here the intensity data points for *q*=0 are ignored, because of the breakdown of the sum rule at *q*→0 limit.

## Supplemental References